\newcommand{\MIG}{{{\textsf{tHinter}}}\xspace}
\documentclass[acmsmall,screen]{acmart}

\AtBeginDocument{%
  }

\setcopyright{acmlicensed}
\copyrightyear{2018}
\acmYear{2018}
\acmDOI{XXXXXXX.XXXXXXX}

\acmConference[Conference acronym 'XX]{Make sure to enter the correct
  conference title from your rights confirmation email}{June 03--05,
  2018}{Woodstock, NY}

\acmISBN{978-1-4503-XXXX-X/18/06}
\usepackage{balance}
\usepackage{graphicx}
\usepackage{caption}
\usepackage{enumitem}
\usepackage{subfigure}
\usepackage{makecell}
\usepackage{multirow}
\usepackage{threeparttable}
\usepackage{xcolor}
\usepackage{listings}
\usepackage{booktabs}
\usepackage{url}
\usepackage{xspace}
\usepackage{algorithm}
\usepackage{amsmath}
\usepackage{algpseudocode}

\begin{document}

\lstset{
    basicstyle=\ttfamily\tiny, % 调整字体大小
    keywordstyle=\color{blue},
    commentstyle=\color{gray},
    stringstyle=\color{red},
    numberstyle=\tiny\color{gray},
    stepnumber=1,
    numbersep=10pt,
    backgroundcolor=\color{white},
    showspaces=false,
    showstringspaces=false,
    showtabs=false,
    frame=single,
    tabsize=2,
    captionpos=b,
    breaklines=true,
    breakatwhitespace=false,
    escapeinside={\%*}{*)}
}
\title{Guided Debugging of Auto-Translated Code Using Differential Testing}
% \title{\MIG: A Differential Testing Approach for Debugging Auto-Translated Code}
\author{Shengnan Wu}
\orcid{0000-0003-1964-313X}
\affiliation{%
  \institution{School of Computer Science, Fudan University}
  \city{Shanghai}
  \country{China}
}
\email{snwu19@fudan.edu.cn}

\author{Xinyu Sun}
\orcid{0009-0008-3965-3240}
\affiliation{%
  \institution{School of Computer Science, Fudan University}
  \city{Shanghai}
  \country{China}
}
\email{sunxy23@m.fudan.edu.cn}

\author{Xin Wang}
\orcid{0000-0002-9405-4485}
\affiliation{%
  \institution{School of Computer Science, Fudan University}
  \city{Shanghai}
  \country{China}
}
\email{xinw@fudan.edu.cn}

\author{Yangfan Zhou}
\orcid{0000-0002-9184-7383}
\affiliation{%
  \institution{School of Computer Science, Fudan University}
  \city{Shanghai}
  \country{China}
}
\email{zyf@fudan.edu.cn}

\renewcommand{\shortauthors}{Wu et al.}

\begin{abstract}
Large Language Models (LLMs) hold great promise in the task of code translation. However, the lack of explainability complicates the identification of the inevitable translation errors. In this paper, we propose \MIG, a debugging tool to locate translation errors in auto-translated code. The core idea of \MIG is that correctly translated, the source and translated code should present the same functionalities, giving the same output for the same input. Hence, lines in the translated code responsible for output differences are possibly translation errors. First, \MIG employs fuzzing to generate diverse test cases that thoroughly explore the translated code. Then, \MIG relies on a heuristic algorithm to pinpoint translation errors from coverage information and differential testing execution results of those test cases. This heuristic algorithm is designed to leverage both the statistics and the expertise of developers. Comprehensive experiments with real code show its effectiveness. It reduces 71\% lines developers need to review during debugging and increases the likelihood of the LLM fixing translation errors in a single query by 59\%. Developers generally consider it satisfactory and helpful.

\end{abstract}



\received{20 February 2007}
\received[revised]{12 March 2009}
\received[accepted]{5 June 2009}

\maketitle

\section{Introduction}
Code translation, {\em i.e.}, translating code from one programming language to another, has long been considered critical yet very challenging.  It is fundamental to many real-life software engineering tasks including legacy code handling~[\citeauthor{DBLP:conf/xpu/Feathers04}~\citeyear{DBLP:conf/xpu/Feathers04}], code migration~[\citeauthor{DBLP:journals/peerjpre/AggarwalSH15}~\citeyear{DBLP:journals/peerjpre/AggarwalSH15}] and code reuse~[\citeauthor{krueger1992software}~\citeyear{krueger1992software}]. 
Recently, large language models~(LLMs) have proven to be promising in code translation~[\citeauthor{DBLP:conf/kbse/HuangWXWCXL23}~\citeyear{DBLP:conf/kbse/HuangWXWCXL23};~\citeauthor{eniser2024towards}~\citeyear{eniser2024towards}]. Besides the impressive performances, simplicity also promotes the adoption of LLMs in code translation. LLMs require only prompt engineering, in contrast to traditional approaches relying on human-constructed rules, supervised algorithms, and code pairs for training. As a result, LLMs are increasingly adopted in code translation. For example, IBM presents the Watsonx Code Assistant\footnote{Watsonx Code Assistant:~\url{https://www.ibm.com/products/watsonx-code-assistant}} to translate Cobol to Java, and a series of commercial LLM based translation tools \footnote{CodeConvert:~\url{https://www.codeconvert.ai/r-to-python-converter}} can be accessed easily via network services.

However, LLMs make translation errors~[\citeauthor{DBLP:journals/pacmse/Yang0YK0LHMJ024}~\citeyear{DBLP:journals/pacmse/Yang0YK0LHMJ024}; \citeauthor{pan2024lost}~\citeyear{pan2024lost}], despite efforts of enhancing the translation quality~
[\citeauthor{DBLP:conf/iui/WeiszMRMHATR22}~\citeyear{DBLP:conf/iui/WeiszMRMHATR22};~\citeauthor{DBLP:conf/iclr/SzafraniecRLLCS23}~\citeyear{DBLP:conf/iclr/SzafraniecRLLCS23}].
This leads to a buggy and non-equivalent code translation. Recent studies show that only 2.1$\%$ to 47.3$\%$ in 1700 code snippets can be correctly translated by the LLM~[\citeauthor{pan2024lost}~\citeyear{pan2024lost}]. Correct code translation relies on finding equivalent API combinations cross language and data type inference, which is inherently hard~[\citeauthor{DBLP:conf/nips/RoziereLCL20}~\citeyear{DBLP:conf/nips/RoziereLCL20}]. 
In addition, the hallucination~[\citeauthor{samek2017explainable}~\citeyear{samek2017explainable}] of LLMs also adds to the difficulties of precise translation. Errors are generally inevitable in code translation by LLMs.

Translation errors can be handled by first pinpointing their locations and then preparing fixes, \textit{i.e.,}  a typical debugging process.
However, debugging has long been considered challenging~[\citeauthor{brooks1974mythical}~\citeyear{brooks1974mythical}]. Debugging auto-translated code by LLMs is even more difficult due to the lack of model explainability. Developers usually have to execute the code several times to collect information about code behavior~[\citeauthor{agrawal1991towards}~\citeyear{agrawal1991towards};~\citeauthor{DBLP:conf/kbse/AugustonJU02}~\citeyear{DBLP:conf/kbse/AugustonJU02}] analyzed by their expertise and experiment with their fixes for several times. This practice is undoubtedly human effort intensive and time-consuming. Hence, an automatic approach to locate translation errors is of critical importance. Nevertheless, designing such an approach is quite difficult. The source and translated codes are with different sets of APIs and grammar, let alone different implementations of the same functionality. So directly comparing the translated code and source by text or AST will be ineffective for locating translating errors.

Even so, the source and translated code should present the same functionalities if correctly translated, giving the same outputs with the same input. This inspires us to locate translation errors by differential testing: lines in the translated code causing output differences are closely related to translation errors. Following this idea of differential testing, two challenges remain unsolved. Firstly, a set of test cases, both valid and comprehensively exploring the behavior of the translated code is needed (challenge 1). If not fully explored, translation error related lines may be missed. Hence, techniques like randomized generation are inapplicable. Secondly, while the fail or pass result of one test case in differential testing indicates whether translation errors exist, it does not pinpoint their exact location. Since the source and translated code are in different programming languages with varying code structures, directly comparing code coverages in the translated code and source code of the same test case is ineffective for localizing translation errors. Therefore, a novel localization algorithm is needed (challenge 2).

To this end, we present \textit{trans-compiler Hinter} (\MIG), an automatic approach to locate translation errors in auto-translated code. The core idea is to find lines causing output differences in the translated code, which are closely related to translation errors.
\MIG first leverages a fuzzing tool \textit{i.e.,} AFL++~[\citeauthor{fioraldi2020afl++}~\citeyear{fioraldi2020afl++}], to generate test cases in a coverage-guided way, which solves the challenge of comprehensive test case generation (challenge 1). Next, \MIG introduces a heuristic-based algorithm to locate translation errors using the execution results of test cases (challenge 2). This algorithm combines two key elements: statistics and developer expertise. Specifically, it calculates the conditional probability of each line being incorrectly translated, based on the coverage information from both passing and failing test cases in differential testing. At the same time, it incorporates insights from developers' past debugging experiences to identify lines more likely to be mistranslated. Finally, \MIG provides those localized lines to the debugger, in the form of a debugging suggestion. This debugging suggestion is expected to facilitate fixing translation errors both manually and by LLMs.

Extensive experiments have demonstrated the effectiveness of \MIG. When developers manually debug translated code without the assistance of other automated tools, \MIG reduces the number of lines they need to examine by an average of 71\%, guiding them to focus on areas where translation errors are more likely to occur. When developers use LLMs to debug translated code, \MIG increases the likelihood of the LLM fixing translation errors in a single query by 59\%. It also enhances the perceived quality of the changes made by the LLM by 13\%. Developers generally consider \MIG to be helpful and effective.

The contributions are summarized below:

\begin{itemize}[leftmargin=*] \item We present an automatic approach to locate translation errors and provide debugging suggestions for auto-translated code. It formulates the task of locating translation errors as a differential testing task. This formulation can shed light on similar future work.

\item We design a novel localization algorithm that identifies translation errors from the execution results of test cases. The algorithm pinpoints errors using conditional probabilities and insights from developers' past debugging experiences. This can serve as a reference for designing other bug localization algorithms.

\item We implement \MIG and will later release it as an open-source tool. Extensive studies using both subjective and objective metrics illustrate its effectiveness. \MIG significantly facilitates debugging by directing both human and LLM attention to areas more likely to contain translation errors. 
% This provides a reference for the evaluation of similar tools. 
\end{itemize}

\section{A Motivating Example}
% \subsection{A motivating example: an incorrectly translated code snippet}
This section presents an incorrectly translated code snippet and preliminary explorations of debugging it. 
 During our dataset construction, which will be elaborated in the evaluation section, a GPT-3.5 based translation tool was implemented to translate Python solutions from LeetCode~\footnote{LeetCode is an online platform offering coding challenges to help developers improve their programming skills. It can be accessed by:\url{https://leetcode.com}} to C++. The Python solution of Question No.5 defines a class \textit{Solution} with two methods, \textit{expand} and \textit{longestPalindrome}. The \textit{longestPalindrome} method uses Manacher's algorithm to find the longest palindromic substring in a given string s. In this method, a \textit{join} function is used to insert a \textit{\#} between every character in the string s to ensure an odd number of length. In the return line, a step parameter is given so the longest palindromic substring can be sliced out. The translated C++ code only presents two major differences, leading to incorrect translation. Firstly, in the \textit{longestPalindrome} method two \textit{\#}s are appended at the beginning and end of the strings, instead of between every character. Secondly, in the return line, no step parameter is used. The Python code and C++ code are shown in Figure~\ref{fig:code-comparison}.
 
We conducted a preliminary debugging experiment and presented the mentioned code pair to five developers (P1-P5). They possessed different levels of proficiency in Python and C++, from beginners to experts. The results revealed the challenging nature of debugging auto-translated code. No participants had successfully fixed the translation errors during the experiment (30 minutes). They all expressed significant difficulty during debugging.

Firstly, one must first understand the Python code before fixing the translation errors. An average of 17 minutes was spent on code understanding. " \textit{I'd rather rewrite it. You have no idea of it unless you read the source code line by line}", stated p5, a medium-level C++ and Python developer. Secondly, pinpointing translation errors based on only the source and translated code is also challenging. "\textit{Function signatures are the same, and structures of the two code snippets look similar. Where can I begin? }" said p1, a beginner in C++. P4, a medium-level Python and C++ developer, noticed the two major differences but failed to recognize them as \textbf{translation errors}. "\textit{No \#s are inserted between characters so no step parameter is used in slicing seems logical. I wouldn't suspect that}. ", said p4. P3, a developer with over 6 years of practice with both C++ and Python, is the only one who fixed the translation errors in the C++ code. "\textit{I just revised the C++ code line by line referring to the Python code.}" said p2. This line-by-line checking and revising process is so time-consuming that p2 took 34 minutes to debug a code snippet under 50 lines. All participants agree that suggestions highlighting potential translation errors would greatly facilitate the debugging process. "\textit{At least I get to know the place to start with, even if the information is imperfect.}" summarized p2, an experienced Python developer and a C++ beginner. 
\begin{figure}[htbp]
    \centering
    \begin{minipage}[t]{0.47\linewidth}
    \centering
    \lstset{language=Python}
    \captionof{lstlisting}{Python Solution}
    \begin{lstlisting}
class Solution:
# Expand around the center and find the maximum length of palindrome
    def expand(self, s, left, right):
        while left >= 0 and right < len(s) and s[left] == s[right]:
            left -= 1
            right += 1
        return (right - left - 2) // 2
        
# Find the longest palindromic substring
    def longestPalindrome(self, s: str) -> str:
        end, start = -1, 0
        s = '#' + '#'.join(list(s)) + '#'
        arm_len = []
        right = -1
        j = -1

        for i in range(len(s)):
            if right > i:
                i_sym = 2 * j - i
                min_arm_len = min(arm_len[i_sym], right - i)
                cur_arm_len = self.expand(s, i - min_arm_len, i + min_arm_len)
            else:
                cur_arm_len = self.expand(s, i, i)
            arm_len.append(cur_arm_len)
            if i + cur_arm_len > right:
                j = i
                right = i + cur_arm_len
            if 2 * cur_arm_len + 1 > end - start:
                start = i - cur_arm_len
                end = i + cur_arm_len

# Remove the special characters and return the longest palindrome
        return s[start+1:end+1:2]
    \end{lstlisting}
    \end{minipage}
    \hfill
    \begin{minipage}[t]{0.47\linewidth}
    \centering
    \lstset{language=C++}
    \captionof{lstlisting}{ Translated C++ Solution}
    \begin{lstlisting}
class Solution {
public:
    int expand(string s, int left, int right) {
        //...basically same as the Python code...;
    }

    string longestPalindrome(string s) {
        int end = -1, start = 0;
        s = "#" + s + "#";
        vector<int> arm_len;
        int right = -1;
        int j = -1;

        for (int i = 0; i < s.size(); ++i) {
            int cur_arm_len;
            if (right > i) {
                int i_sym = 2 * j - i;
                int min_arm_len = min(arm_len[i_sym], right - i);
                cur_arm_len = expand(s, i - min_arm_len, i + min_arm_len);
            } else {
                cur_arm_len = expand(s, i, i);
            }
            arm_len.push_back(cur_arm_len);
            if (i + cur_arm_len > right) {
                j = i;
                right = i + cur_arm_len;
            }
            if (2 * cur_arm_len + 1 > end - start) {
                start = i - cur_arm_len;
                end = i + cur_arm_len;
            }
        }
        return s.substr(start + 1, end - start);
    }
};
    \end{lstlisting}
    \end{minipage}
    \hfill
    \caption{A Python Solution of \textit{Longest Palindrome Substring} and the Auto-translated C++ Version}
    \label{fig:code-comparison}
\end{figure}

Findings from the preliminary debugging experiment emphasize the importance of an debugging suggestion that highlights potential translation errors. Moreover, 
our exploration with Stack Overflow~\footnote{Stack Overflow is a popular Q\&A platform where developers can ask and answer coding-related questions, sharing knowledge across the programming community. It can be accessed by: ~\url{https://stackoverflow.com}} illustrates that relying on manual replies from others is not a reliable way to obtain such suggestions.

We searched with the tag "[code-translation]", and found 387 discussions about debugging auto-translated code till 28th, May, 2024. Among them, 36 discussions received zero replies and 71 did not receive satisfactory replies (indicated by the adoption of who posted the discussion). This indicates around one-third of incorrectly translated code did not receive effective debugging suggestions. Moreover, a frequent reply was advising "\textit{not translating but rewriting the code in another language}", for "\textit{those translating tools are often unreliable}". Usually, this advice would be responded to by stating unfamiliarity with the target programming language. This confirms our observation that developers who are not experts in both source-translated programming languages are more motivated to use translation tools. Hence, an automatic method to locate translation errors and provide a debugging suggestion accordingly is of crucial concern.

\section{Methodology}
\begin{figure*}[ht!]
    \centering
    \includegraphics[width=0.95\linewidth]{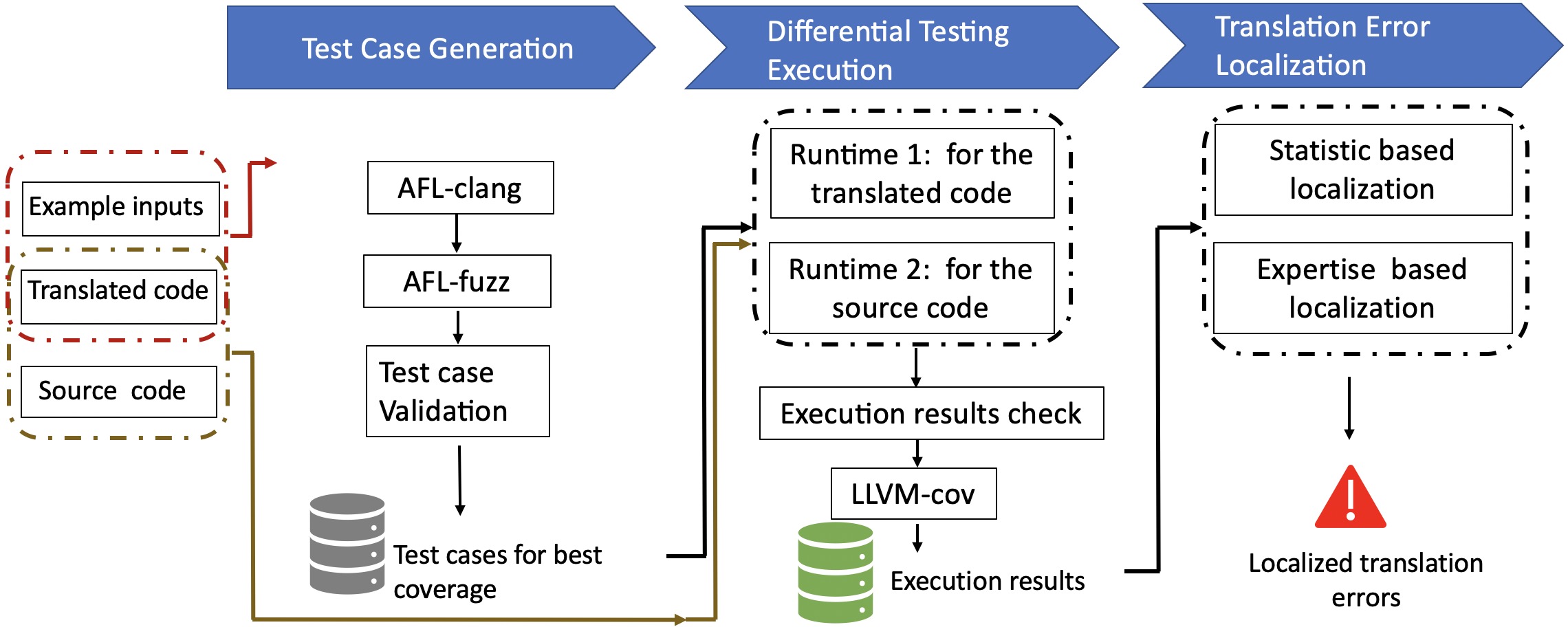}
    \vspace{-8pt}
    \caption{Overview of \MIG}
    \label{fig:overview}
    \vspace{-8pt}
\end{figure*}

This section presents the technique design of our proposed approach \MIG. \MIG locates translation errors and gives debugging suggestions accordingly. The core idea is to find lines in the translated code causing output differences from the source code. Those lines are closely related to translation errors, as they incur altered functionalities. \MIG  takes the source code, translated code, and seeds, \textit{i.e.,} example inputs of the translated code as input and the localized lines, in the form of a debugging suggestion, as the output.
The overall workflow of \MIG is illustrated in Figure ~\ref{fig:overview}. \MIG includes three steps: 1) fuzzing based test case generation, 2) differential testing execution, and 3) translation error localization.

In the fuzzing based test case generation step, \MIG first generates test cases using coverage-guided fuzzing and then excludes invalid cases with a rule-based filter. This ensures that the test cases are valid and can comprehensively explore the behavior of the translated code. In the differential testing execution step, \MIG conducts differential testing and collects execution results of test cases, including coverage information for each test case. In the translation error localization step, \MIG leverages a localization algorithm, designed to incorporate both statistics and developers' expertise accumulated from past debugging experiences, to locate translation errors. 
\subsection{Test Case Generation}
In this step, \MIG generates test cases according to the translated code and seeds. These test cases should 1) explore the behavior of the translated code comprehensively and 2) be valid for the translated code. To explore the behavior of the translated code comprehensively, we leverage the fuzzing tool AFL++. Specifically, we first conduct instrumentation by AFL-clang so \MIG could carry out a coverage-guided generation process. Then AFL-fuzz will keep mutating the seeds until the stop condition. An ideal condition would be that the generated test cases reach 100\% line coverage. However, due to the existence of lines like header files, which will not be counted in the coverage calculation, 100\% line coverage would be impractical. So we set the stop condition to 90\% line coverage. Due to consideration of quick response, for corner cases where 90\% are also time-consuming to achieve out of high complexity and huge exploring space or dead codes, AFL++ explores the code for one minute. 

To guarantee the validity of the generated test cases, we validate them by a rule-based filter. Specifically, due to the consideration of maximizing coverage, the mutation of AFL-fuzz may incur special characters like ASCII in the test cases. This may cause validity issues to the translated code. This filer excludes special characters like ASCII, punctuation marks, and characters from minor natural languages. By our estimation, generally only less than 20 percent of generated test cases are excluded, so the coverage will not be compromised. 
\subsection{Differential Testing Execution}
% change information collection to differential testing conduction
 This step conducts differential testing and collects information needed in the localization of translation errors. Besides the pass or fail test results of test cases, line coverage of each test case is also collected. Firstly, we adopt two runtimes for the source and translated code to execute all test cases accessed from the test case generation step.  
Pass or fail of a test case is determined with the following oracle, which is essentially a consistency check. 
Given test cases \( X \) for a < source, translated > code pair $t$, the function \( S(X) \) represents the execution result /output of the source code induced by $X$. The function \( T(X) \) represents the execution result/output of the translated code induced by $X$. Formally, for any \( X \), we define value of the test result \( y \) as follows:
\begin{equation}
y = \begin{cases} 
\text{Pass} & \text{if } S(X) == T(X), \\
\text{Fail} & \text{if } S(X) != T(X).
\end{cases}
\end{equation}
 As instrumentation is conducted before compiling the translated code, execution results and coverage information will be logged during execution. Next we parse the log by LLVM-cov to obtain which lines are covered by each test case in the translated code. By this way, we get all the information needed for translation error localization. Examples of execution results, including line coverages, are shown in Table~\ref{table:dataexample}. 
 \begin{table}[h!]
\centering
\caption{Examples of Execution Results.}
\label{table:dataexample}
\begin{tabular}{l|c|c|c|c}
\hline
\textbf{Test case id} & \textbf{\makecell{Lines covered \\ in translated code}} & \textbf{\makecell{Execution results \\ of source code}} & \textbf{\makecell{Execution results \\ in translated code}} & \textbf{Testing Results} \\
\hline
3421 & 2, 7, 31, 62 & abca & abcad & Fail \\
% \hline
3422 & 5, 7, 31, 62 & abcae & abcae & Pass \\
\hline
\end{tabular}
\end{table}

 The number of test cases executed affects the performance of \MIG. Less test cases executed decrease response time and lower the quality of generated suggestions. More test cases work oppositely. We leave this trade-off to users. By default \MIG executes all test cases generated. Users can set it to a given number if a faster response is desirable in a specific usage context.
\subsection{Translation Error Localization}
In this step, \MIG localizes translation errors from execution results of test cases. A heuristic-based localization algorithm, leveraging statistics and the expertise of developers is designed and used for locating translation errors. Specifically, it calculates the conditional probability of each line being incorrectly translated, based
on the coverage information from both passing and failing test cases in differential testing. At the
same time, it refers to insights from developers’ past debugging experiences to identify lines
more likely to be mistranslated. Finally, it provides those localized lines, in the form of a debugging suggestion.

From the perspective of statistics, if coverage of certain lines in the translated code is strongly related to a \textbf{Fail} testing result, \textit{i.e.,} different outputs from the source and translated code, these lines are possibly incorrectly translated. Hence we obtain two core heuristics based on statistics:
\begin{itemize}
    \item \textbf{Heuristic 1:} If test cases covering certain lines in the translated code all achieve the \textbf{Fail} testing result, those lines are highly likely to be incorrectly translated.
    % \item \textbf{Heuristic 2:} If the posterior possibility of covering a specific line in the translated code and reaching \textbf{Fail} exceeds the threshold of a random assumption (50\%), this line is likely to be incorrectly translated.
    \item \textbf{Heuristic 2:} If covering a specific line in the translated code and reaching a \textbf{Fail} test result occurs more frequently than a random assumption (50\%), this line is likely to be incorrectly translated.
\end{itemize}

Those two heuristics inspire us to calculate the conditional possibilities of each line being incorrectly translated, based on coverage as well as pass or fail of each test case in the differential testing. The two heuristics are further summarized into a \textit{Localize by Statistics} component, which is demonstrated with pseudo-code in Algorithm~\ref{statistics}.  A \textit{suspicious score by statistics} is assigned to each line according to the calculated conditional possibilities, indicating how likely this line is incorrectly translated. The $\theta_{punish}$ adds different weights to situations in Heuristic 1 and 2. A $baseScore$ is added only for regularization reasons, avoiding negative values. 
\begin{algorithm}
\caption{Localize by Statistics}
\label{statistics}
\begin{algorithmic}
\For{each $line \in \texttt{translatedCode}$}
    \State $passedCases, failedCases, allCases\gets \{\}$
    \For{each $case \in \texttt{findCasesThatCovered}(line)$}
        \If{\texttt{isCasePassed}(case)}
            \State $passedCases \gets passedCases \cup \{ case \}$
        \Else
            \State $failedCases \gets failedCases \cup \{ case \}$
        \EndIf
    \EndFor
    \State $suspiciousScoreByStatistics = baseScore + P(\frac{n_{p}}{n_{f}}) * \theta_{punish}$

\EndFor
\end{algorithmic}
\end{algorithm}

Besides statistics, \MIG also leverages the expertise of developers in the localization algorithm. Past debugging experiences, once internalized and processed through the developers' expertise, often lead to the empirical conclusion that certain parts of the code are more prone to translation errors. This empirical conclusion prioritizes checking the parts that are more prone to translation errors and thus contributes to the localization of translation errors. To capture such an empirical conclusion and extract it into helpful heuristics, we conducted a 40-minute round table discussion with three developers about their past debugging experiences. These developers represent varying levels of expertise—experienced, medium, and novice—which allows us to draw more generally applicable insights. Through the round table discussion, we identified other three heuristics.
\begin{itemize}
\item \textbf{Heuristic 3:} Control flow-related code, such as conditional statements and loops, is more prone to translation errors.
\item \textbf{Heuristic 4:} Code that itself is of a specific type is more likely to be translated incorrectly compared to code within the scope of that specific type.
\item \textbf{Heuristic 5:} Structurally simple code (without scope) is less prone to translation errors.
\end{itemize}

Those three heuristics are summarized into a \textit{Localize by Expertise} component, which is shown with pseudo-code in Algorithm~\ref{expertise}. A \textit{suspicious score by expertise} is assigned to each line of the translated code, indicating how prone to translation errors each line is according to the past debugging experience of developers. Different values of $\alpha$ are used in this \textit{Localize by Expertise} component, assigning different weights to different heuristics. It is worth noting that we adjust weights in the \textit{Localize by Expertise} component by both adding and multiplying operations, while in the \textit{Localize by Statistics} component, only multiplying operations are used. So the localization \MIG conducts are mainly based on statistics, being more objective.
\begin{algorithm}
\caption{Localize by Expertise}
\label{expertise}
\begin{algorithmic}
\For{each $line \in \texttt{translatedCode}$}
    \State $susceptibleSyntaxUnit \gets set$
    \State $suspiciousScoreByExpertise \gets \texttt{getScore}(line)$
    \For{each $case \in \texttt{findCasesThatCovered}(line)$}
        \State $syntaxUnit, syntaxScope \gets \texttt{analyzeSyntaxByLine}$
        \If{$syntaxUnit \in susceptibleSyntaxUnit$}
            \State $suspiciousScoreByExpertise = suspiciousScore + \alpha_1$
         \EndIf
        \If{$syntaxScope \in susceptibleSyntaxUnit$}
            \State $suspiciousScoreByExpertise = suspiciousScoreByExpertise * \alpha_2$
        \EndIf
    \EndFor
\EndFor
\end{algorithmic}
\end{algorithm}

\MIG conducts translation error localization via identifying lines with significantly higher \textit{suspicious scores.} Specifically, an \textit{overall suspicious score} is calculated for each line, adding the \textit{suspicious score by statistics} and \textit{suspicious score by expertise}. Lines exceeding a given threshold are flagged as potential translation errors. If no lines exceed the threshold, a simple anomaly detection, two standard divisions above the mean, is carried out to identify lines with comparatively higher \textit{overall suspicious scores}.  

\section{Evaluation}
In this section we evaluate \MIG's performance by answering the following research questions (RQs):
\begin{itemize}
    \item \textbf{RQ1:} What's the performance of \MIG in manual debugging?
    \item \textbf{RQ2:} What's the performance of \MIG in debugging with LLMs?
    \item \textbf{RQ3:} How do the numbers of test cases executed affect the performance of \MIG?
\end{itemize}
RQ1 aims to illustrate how can \MIG help with debugging by manual efforts of developers, without the help of other automatic tools. RQ2 aims to unveil how can \MIG help with fixing translation errors by LLMs. After exploring two commonly seen debugging settings (manually and by LLMs), we target \MIG's key component. RQ3 presents how different numbers of test cases executed affect the performance.
\subsection{Experiment Setup}
% dataset, metrics, experiment environment
\subsubsection{Data Collection}\hspace*{\fill}\newline
\textbf{A translation tool for data collection.}
We focus on the Python to C++ scenario during the evaluation of \MIG. Although the method itself is applicable across many programming languages, we choose this scenario for the comparatively significant differences between the two languages' grammar. To prevent overfitting to a specific translation tool, we implement a generalized LLM-based one to construct datasets for evaluation. This translation tool only serves as a provider of <source, translated> code pairs. Considering the objectivity and fairness of our evaluation with \MIG, this translation tool should not present poorly translated code pairs and at the same time avoid techniques not generally applicable. To this end, we choose GPT-3.5 as the base model. In-context learning~[\citeauthor{min2022rethinking}~\citeyear{min2022rethinking}] and human-in-the-loop~[\citeauthor{ge2023openagi}~\citeyear{ge2023openagi}] strategy are used in prompt engineering. 

\begin{figure*}[ht!]
    \centering
    \includegraphics[width=0.95\linewidth]{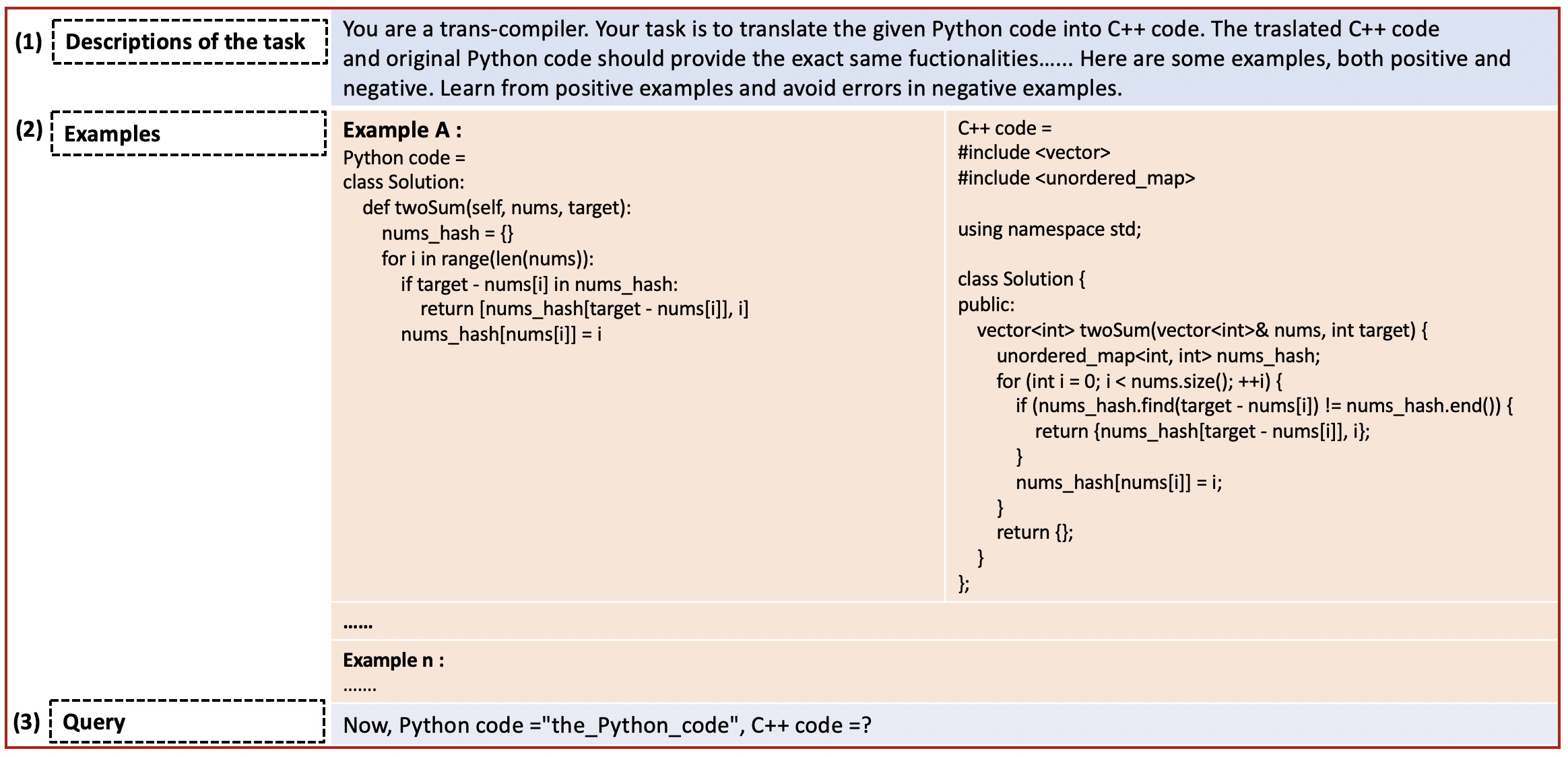}
    \vspace{-8pt}
    \caption{A simplified demonstration of the prompt used in the translation step.}
    \label{fig:translateprompt}
    \vspace{-8pt}
\end{figure*}

\begin{figure*}[ht!]
    \centering
    \includegraphics[width=0.95\linewidth]{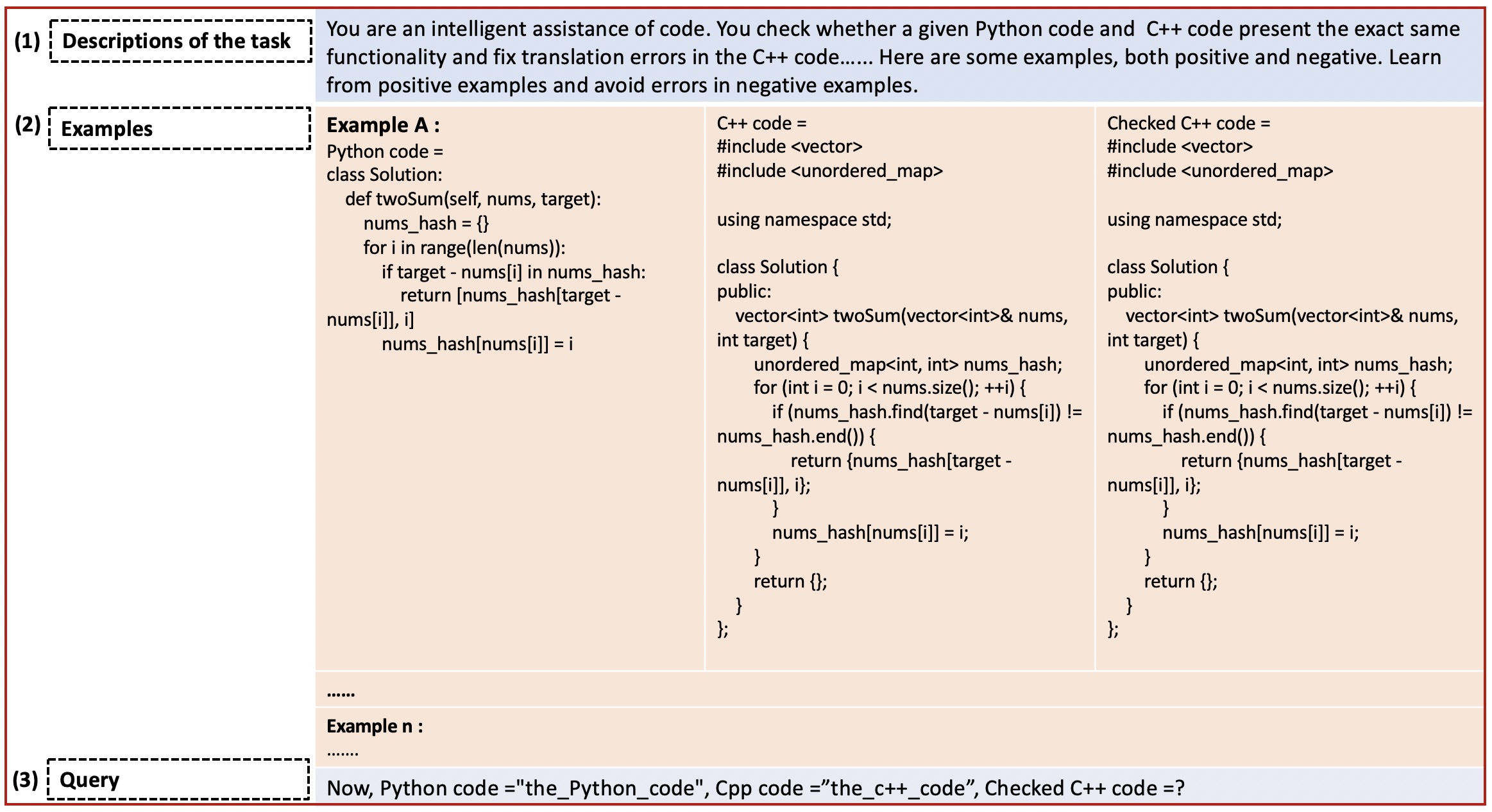}
    \vspace{-8pt}
    \caption{A simplified demonstration of the prompt used in the validation step.}
    \label{fig:checkprompt}
    \vspace{-8pt}
\end{figure*}

Specifically, this translation tool consists of a translation step and a validation step. 
The translation step translates Python code to C++ code. The prompt used in this step consists of a paragraph describing the task and several examples, aiming to provide domain knowledge. A simplified demonstration of the prompt used in the translation step is shown in Figure~\ref{fig:translateprompt}. 
The validation step fixes translation errors in the translated C++ code if there are any. We design this validation step based on the assumption that querying the LLM repeatedly decreases the possibility of unreliable outputs. The prompt used is constructed similarly to that in the translation step.
Several examples, basically tuples of source code, translated code, and fixed code, are provided in the prompt, following a description of the task. A simplified demonstration of the prompt used in the validation step is shown in Figure~\ref{fig:checkprompt}. Initially, examples contained in the prompts are randomly sampled from the official solutions of LeetCode and manually translated. To ensure the performance of our translation tool, we iteratively improve the prompts according to human feedback on the quality of the translations/validation. During this process, examples can be substituted, deleted, or added. We stop the iteration process until no obvious patterns of errors are found in randomly sampled 20 code pairs. 

\textbf{Datasets.}
We use the official Python solutions of LeetCode as the input of our translation tool. A spider is implemented for the purpose of data access. After 2 weeks of interaction with LeetCode's API, we get all Python solutions to 611 questions. Then they are translated into C++ by our translation tool and posted back to the LeetCode China webpage. By this, we can run the official test cases to determine the correctness of the translated code. 1/3 of the translated C++ solutions fail the corresponding test cases, indicating translation errors, in which compiler errors dominate. 

We aim to evaluate \MIG across a wide range of translation errors, particularly those causing runtime errors, where the code runs but its functionality is altered. To achieve this, we adopt a purposive sampling approach~[\citeauthor{campbell2020purposive}~\citeyear{campbell2020purposive}]. We start by constructing a pool of <Python solution, translated C++ solution> pairs that exhibit runtime errors. Then, we identify the corresponding questions associated with these runtime-error solutions. Since there are typically 2-3 Python solutions per question, we expand the pool by including all Python solutions and their translated C++ counterparts for those questions. This ensures the pool covers translation errors that lead to runtime errors, as well as other conditions such as compiler errors, overflows, and correct translations. In total, the pool comprises 211 <Python solution, translated C++ solution> pairs.

The complexity of the translated code may affect the performance of debugging tools like \MIG. Hence, we categorize those 211 code pairs into different complexity levels. We use Complexity Score ($S_{complex}$) to describe the complexity of translated C++ code. Three factors are considered: the difficulty level of the question defined by LeetCode ($S_{difficulty}$), the acceptance rate of the Python solution ($R_{accept}$), and the cyclomatic complexity~[\citeauthor{DBLP:journals/software/EbertC16}~\citeyear{DBLP:journals/software/EbertC16}] of the translated C++ solution ($C_{cyc}$). $S_{complex}$ is defined as: 
\begin{equation}
    S_{complex}=S_{difficulty}*0.33 + R_{accept}*0.33 + C_{cyc}*0.33
\end{equation}
 A $S_{complex}$ typically falls within 0 to 10 and is further mapped to a Complexity Level with:
\begin{equation}
Complexity \ Level =
\begin{cases} 
\text{Low}, & \text{if } S_{complex} < 4 \\
\text{Medium}, & \text{if } 4 \leq S_{complex} \leq 7 \\
\text{High}, & \text{if } S_{complex} > 7
\end{cases}
\end{equation}
Two datasets are constructed from those 211 code pairs and used for the evaluation of \MIG. 
\begin{itemize}
    \item \textbf{Dataset A: } All 211 <Python solution, translated C++ solution> code pairs. The Python solutions in those code pairs should be precise and correct, as they are official solutions from LeetCode. The C++ solutions may contain translation errors. There are 76 \textit{High} complexity pairs, 102 \textit{Medium} complexity pairs, and 36 \textit{Low} complexity pairs. This dataset is used to answer RQ1 and RQ3. 
    \item \textbf{Dataset B: } Randomly sampled 30 <Python solution, translated C++ solution> code pairs from dataset A. There are 10 pairs for each Complexity Level. This dataset is used to answer RQ2.
\end{itemize}
For RQ1 and RQ3, to provide comprehensive and objective conclusions about the performance of \MIG, we use a dataset sufficiently large and includes a full range of complexity levels. RQ2, on the other hand, focuses more on the quality of code changes made by the LLM. Since this evaluation relies more on qualitative analysis and manual assessment, a smaller dataset with evenly sampled complexity levels is more appropriate.

\subsubsection{Metrics}\hspace*{\fill}\newline
We use Reduction Ratio ($R_{reduc}$), Perceived Helpfulness ($H_{perceived}$), Fix Ratio ($R_{fix}$), Attempt Ratio ($R_{attp}$), and Perceived Fix Quality ($Q_{fix}$) to measure the performance of \MIG.

$R_{reduc}$, $R_{fix}$, and $R_{attp}$ describe the performance of \MIG from the \textbf{objective} side. $R_{reduc}$ measures the ratio of lines excluded by \MIG, and is calculated as: \begin{equation}
  R_{reduc}=1-\frac{Len(localized)}{Len(translated)}  
\end{equation}
where $Len(localized)$ and $Len(translated)$ indicate how many lines are in the debugging suggestion \MIG gives and the translated code respectively. Higher $R_{reduc}$ indicates a stronger ability to exclude possibly correctly translated lines during debugging. Thus developers avoid examining the translated code line by line and the debugging efficiency gets enhanced. Lower $R_{reduc}$ works oppositely. $R_{fix}$ measures the possibility of translation errors in the translated code being fixed by the LLM in a single query. It is calculated as:\begin{equation}
  R_{fix}=\frac{N_{fixed}}{N_{all}}  
\end{equation}
where $N_{fixed}$ and $N_{all}$ indicate the number of code snippets the LLM fixes and it should fix respectively. Higher $R_{fix}$ represents a stronger ability to fix translation errors. Lower $R_{fix}$ works oppositely. $R_{attp}$ measures the possibility of the LLM finding translation errors in the translated code and attempting to fix them in a single query. It is calculated as:\begin{equation}
  R_{attp}=\frac{N_{attp}}{N_{all}}  
\end{equation}
where $N_{attp}$ indicates the number of code snippets the LLM attempts to fix. The LLM only fixes code snippets when it realizes there are translation errors. Hence a higher $R_{attp}$ indicates a higher possibility that a translated code gets fixed by the LLM. A lower $R_{attp}$ works oppositely. $R_{reduc}$ is used in RQ1 and RQ3. $R_{fix}$ and $R_{attp}$ are used in RQ2.

$H_{perceived}$ and $Q_{fix}$ describe the performance of \MIG from the \textbf{subjective} side. $H_{perceived}$ measures the perceived helpfulness of \MIG from the perspective of debugging developers on a 1 to 5 scale, with 1 representing merely helpful and 5 strongly helpful. Higher $H_{perceived}$ indicates a higher sense of helpfulness and satisfaction. Lower $H_{perceived}$ works oppositely. $Q_{fix}$ measures the subjectively assessed quality of changes made to the translated code during LLM based debugging. It is also on a 1 to 5 scale, with 1 representing extremely poor quality and 5 extremely high quality. Higher $Q_{fix}$ indicates a stronger ability to facilitate the LLM debugging process. Lower $Q_{fix}$ works oppositely. $H_{perceived}$ is used in RQ1 and RQ3. $Q_{fix}$ is used in RQ2.

\subsubsection{Manual Assessing Procedure}\hspace*{\fill}\newline 
Two subjective metrics, $H_{perceived}$ and $Q_{fix}$, require manual assessment, which is prone to the subjectivity of the conductors. To guarantee validity, a guideline mapping the sense of helpfulness to detailed standards is discussed and shown in Table~\ref{tab:assessment_standards}. Scores of $H_{perceived}$ must be given largely based on this guideline. In addition, if a code snippet is correctly translated and \MIG barely presents valid information, a higher score of $H_{perceived}$ should also be given. For $Q_{fix}$, scores must be given considering those three factors: 1) preserving the functionalities of the source code; 2) fixing the translation errors; and 3) not inducing new errors. The corresponding standards are shown in ~\ref{tab:assessment_standards_quality}.
\begin{table}[h!]
\centering
\caption{Standards Used to Guide the Manual Assessing Process of $H_{perceived}$.}
\label{tab:assessment_standards}
\begin{tabular}{l|c|c}
\hline
\textbf{Scores} & \textbf{Representations} & \textbf{Explanations} \\ 
\hline
\makecell{1} & \makecell{Merely helpful} & \makecell{The localized lines \textbf{are irrelevant to}  translation errors.} \\ 
% \hline
\makecell{2} & \makecell{Somewhat helpful} & \makecell{The localized lines are \textbf{loosely related to} translation errors.} \\ 
% \hline
\makecell{3} & \makecell{Helpful} & \makecell{The localized lines are \textbf{closely related to} translation errors.} \\ 
\makecell{4} & \makecell{Quite helpful} & \makecell{The localized lines \textbf{highlight symptoms} of translation errors.} \\ 
\makecell{5} & \makecell{Strongly helpful} & \makecell{The localized lines \textbf{directly present} translation errors.} \\ 
\hline
\end{tabular}
\end{table}

During the manual assessment, two authors, C1 and C2, act as conductors. C1 is an experienced C++ and Python developer with over three years of industrial experience, while C2 is a mid-level Python developer with basic C++ knowledge. C1's expertise enables a thorough and proficient assessment, whereas C2 represents the perspective of a general translation tool user. Two conductors independently score all code pairs. If significant differences in their scores arise for a single code pair, the reasons are discussed and reviewed. This cross-validation process helps mitigate the impact of human subjectivity.
\begin{table}[h!]
\centering
\caption{Standards Used to Guide the Manual Assessing Process of $Q_{fix}$.}
\label{tab:assessment_standards_quality}
\begin{tabular}{l|c|c}
\hline
\textbf{Scores} & \textbf{Representations} & \textbf{Explanations} \\ 
\hline
\makecell{1} & \makecell{Extremely poor} & \makecell{Changes \textbf{do not fix} translation errors, incur \textbf{new ones} and \\ include \textbf{unnecessary paraphrasing}.}\\ 
% \hline
\makecell{2} & \makecell{Poor} & \makecell{Changes \textbf{do not fix} translation errors and include \\ \textbf{unnecessary paraphrasing}.}\\ 
% \hline
\makecell{3} & \makecell{Fair} & \makecell{Changes are essentially \textbf{paraphrases} of the old code.\\  }\\ 
\makecell{4} & \makecell{Good} & \makecell{Changes \textbf{fix} translation errors but contain\\\textbf{unnecessary paraphrasing}.} \\ 
\makecell{5} & \makecell{Extremely high} &\makecell{Changes \textbf{fix} translation errors, incur \textbf{no new ones} and\\  do not include \textbf{unnecessary paraphrasing}.} \\ 
\hline
\end{tabular}
\end{table}
\subsubsection{Experimental Environment}\hspace*{\fill}\newline
We implement \MIG with 1741 lines of Python code. 
All experiments are conducted on a MacOS Sonoma 14.5 system with an Apple M2 Max chip, and 32GB RAM.
\subsection{Performance in Manual Debugging (RQ1)}
\begin{figure*}[ht]
    \centering
    \subfigure[overview of $R_{reduc}$.]{
        \begin{minipage}[t]{0.3\linewidth}
        \centering
        \includegraphics[width=\linewidth]{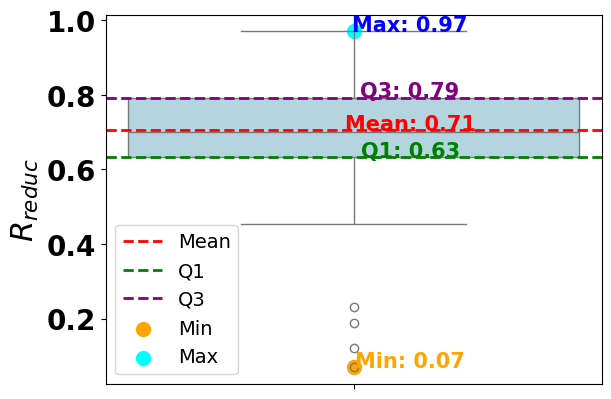}
        \vspace{-4pt}
        \label{fig:figure1}
        \end{minipage}
    }
    \subfigure[Complexity Level Distributions.]{
        \begin{minipage}[t]{0.3\linewidth}
        \centering
        \includegraphics[width=\linewidth]{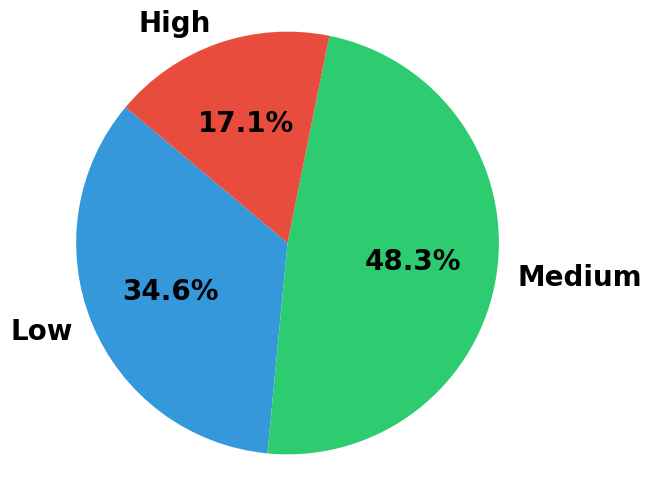}
        \vspace{-4pt}
        \label{fig:figure2}
        \end{minipage}
    }
    \subfigure[$R_{reduc}$ by Complexity Levels.]{
        \begin{minipage}[t]{0.3\linewidth}
        \centering
        \includegraphics[width=\linewidth]{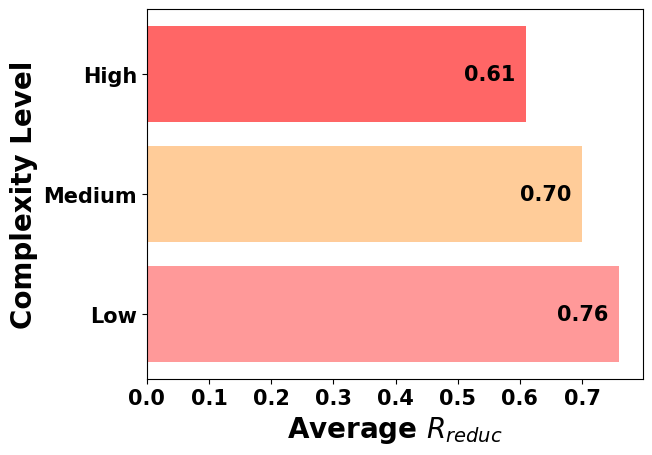}
        \vspace{-4pt}
        \label{fig:figure3}
        \end{minipage}
    }
    \vspace{-12pt}
    \caption{\MIG's Performance Evaluated by $R_{reduc}$.}
    \label{fig:combined_figures}
    \vspace{-8pt}
\end{figure*}

\MIG significantly reduces the number of lines developers need to check when debugging translated code. As shown in Figure~\ref{fig:figure1}, the average $R_{reduc}$ is 0.71, meaning that, on average, 71\% of the code no longer requires manual inspection. Furthermore, for 50\% of the samples, 63\% to 79\% of the lines in the translated code do not need to be examined line by line, with the first quartile ($Q1$) of $R_{reduc}$ being 0.63 and the third quartile ($Q3$) being 0.79. 

\MIG reduces more lines for \textit{Low} complexity samples. The distribution of complexity for all samples is shown in Figure~\ref{fig:combined_figures2}. Around 2 thirds (65.5\%) of samples are in the \textit{High} complexity group and \textit{Medium} complexity group. This is consistent with the construction of dataset A. More complex code may be more easily incorrectly translated, leading to runtime errors. 
\MIG achieves the highest $R_{reduc}$ in the \textit{Low} complexity group. One possible reason is that less complex code may have a simpler structure, like fewer control flows. In this way a single test case could cover more lines in the translated code, resulting in more excluded lines and a higher $R_{reduc}$.

\MIG presents an average of 3.18 in $H_{perceived}$, the place between \textbf{Helpful} and \textbf{Quite helpful}. This indicates developers are generally satisfied and consider \MIG effective. $H_{perceived}$ for each complexity group remains stable, indicating stable performance due to a generally applicable method. However, C2 generally gives higher scores than C1. We find that is caused by a higher sense of helpfulness of \MIG for non-experts of both source and translated programming languages. C2, representing general users of translation tools, only has basic knowledge of C++. We find for C2 the most challenging part of debugging auto-translated code is determining where to start the examination. \MIG guides C2 to examine lines more possibly containing translation errors. Hence, a higher sense of helpfulness is achieved. Even when the suggestions by \MIG are merely/somewhat helpful, they still avoid C2 from line by line examination for lack of practice in C++.

\begin{table}[h!]
\centering
\caption{$H_{perceived}$ of \MIG by two conductors (C1 and C2) across different complexity levels.}
\label{tab:helpfulness}  % 添加标签
\begin{tabular}{l|c|c|c|c}
\toprule
        \textbf{Conductors}  & \textbf{Low~$\uparrow$} & \textbf{Medium~$\uparrow$} & \textbf{High~$\uparrow$} & \textbf{Average~$\uparrow$} \\
\midrule
\textbf{C1}       & 3.10          & 3.10              & 3.06         & 3.09              \\
\textbf{C2}       & \textbf{3.25}           & \textbf{3.26}             & \textbf{3.28}           & \textbf{3.26}               \\
\textbf{Overall}  & 3.17           & 3.18              & 3.17           & 3.18             \\
\bottomrule
\end{tabular}
\end{table}

%加上方框总结结论
\begin{center}
\fbox{
    \parbox{0.9\linewidth}{\textbf{Summary:} 
    \MIG significantly reduces lines developers check when debugging translated code. With \MIG, an average of 71 \% of lines no longer require manual inspection. Developers generally consider \MIG satisfactory and helpful during debugging. It offers a larger sense of helpfulness to non-experts for \MIG guides them to examine where translation errors emerge more possibly.
    % 解释原因
    }
}
\end{center}

\subsection{Performance in Debugging with LLMs (RQ2)}
\textbf{Baselines.} We unveil the ability of \MIG by contrasting debugging auto-translated code via the LLM with/without \MIG. Two prompts are constructed based on GPT-3.5. 
\begin{itemize}
    \item \textbf{With \MIG:} The source code, translated code, localized translation errors by \MIG, and a paragraph describing this debugging task are all included in the prompt. We use a standard zero-shot strategy, avoiding complex prompt engineering techniques affecting the results.
    \item \textbf{Without \MIG:} Only the source code, translated code, and description of the debugging task are included in the prompt.  
\end{itemize}

\begin{figure*}[ht]
    \centering
    \subfigure[Comparing of average $R_{fix}$ with/without \MIG.]{
        \begin{minipage}[t]{0.45\linewidth}
        \centering
        \includegraphics[width=\linewidth]{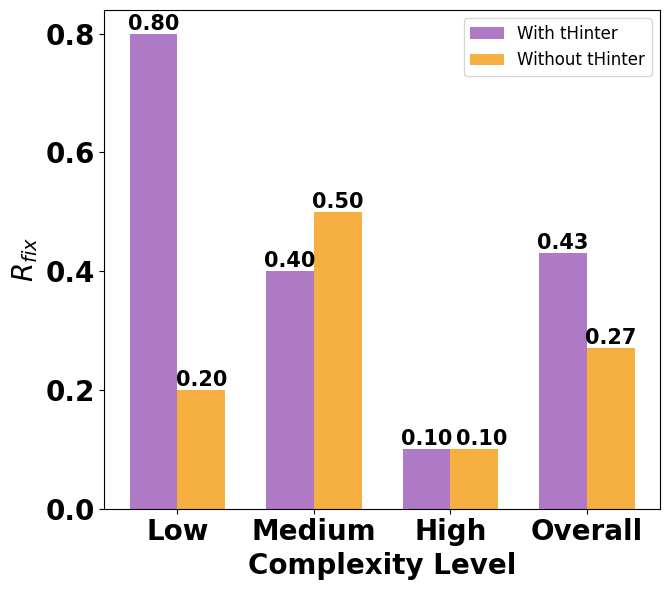}
        \vspace{-4pt}
        \label{fig:figure4}
        \end{minipage}
    }
    \subfigure[Comparing of average $R_{attp}$ with/without \MIG.]{
        \begin{minipage}[t]{0.45\linewidth}
        \centering
        \includegraphics[width=\linewidth]{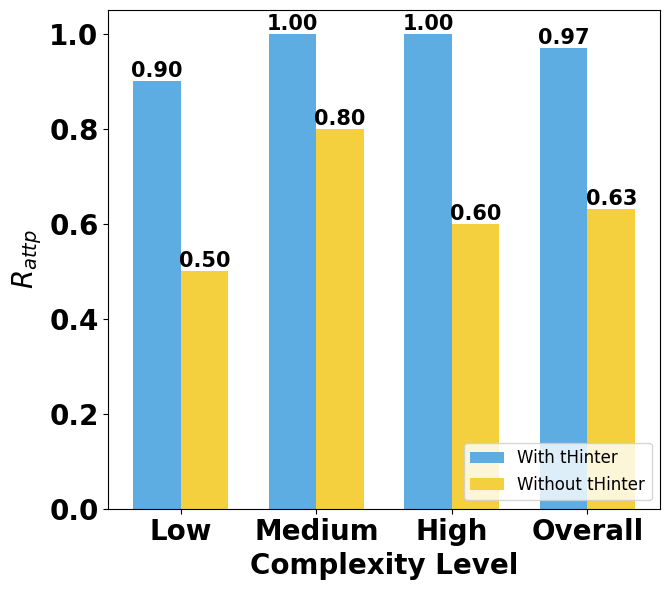}
        \vspace{-4pt}
        \label{fig:figure5}
        \end{minipage}
    }
    
    \vspace{-12pt}
    \caption{ Fixing translation errors by the LLM with/without \MIG.}
    \label{fig:combined_figures2}
    \vspace{-8pt}
\end{figure*}

\textbf{Results.} \MIG increases the possibility of translation errors being fixed by the LLM during a single query by 59\%. According to Figure~\ref{fig:figure4}, with \MIG, the $R_{fix}$ overall achieves 0.43, which is 59\% higher than that without \MIG. This performance advantage can be partially explained with $R_{attp}$. As shown in Figure~\ref{fig:figure5}, with \MIG, $R_{attp}$s are significantly accelerated, overall and across different complexity groups. This indicates that potential translation errors localized by \MIG, essentially context information, help the LLM to understand the existence of translation errors and hence trigger the fixing attempts~[\citeauthor{pan2024lost}~\citeyear{pan2024lost}]. Without this, the LLM judges some incorrectly translated code snippets as correctly translated and stops the fixing attempts, eventually leading to a lower $R_{fix}$.

\begin{figure*}[ht]
    \centering
    \subfigure[Overview of average $Q_{fix}$ with/without \MIG.]{
        \begin{minipage}[t]{0.45\linewidth}
        \centering
        \includegraphics[width=\linewidth]{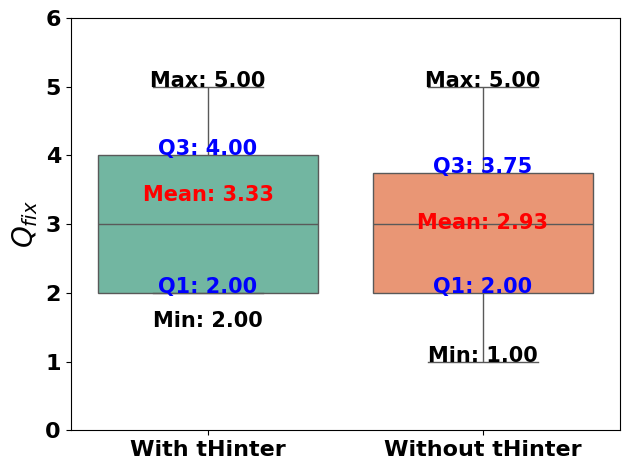}
        \vspace{-4pt}
        \label{fig:figure6}
        \end{minipage}
    }
    \subfigure[Comparing of average $Q_{fix}$ with/without \MIG within different Complexity Levels.]{
        \begin{minipage}[t]{0.50\linewidth}
        \centering
        \includegraphics[width=\linewidth]{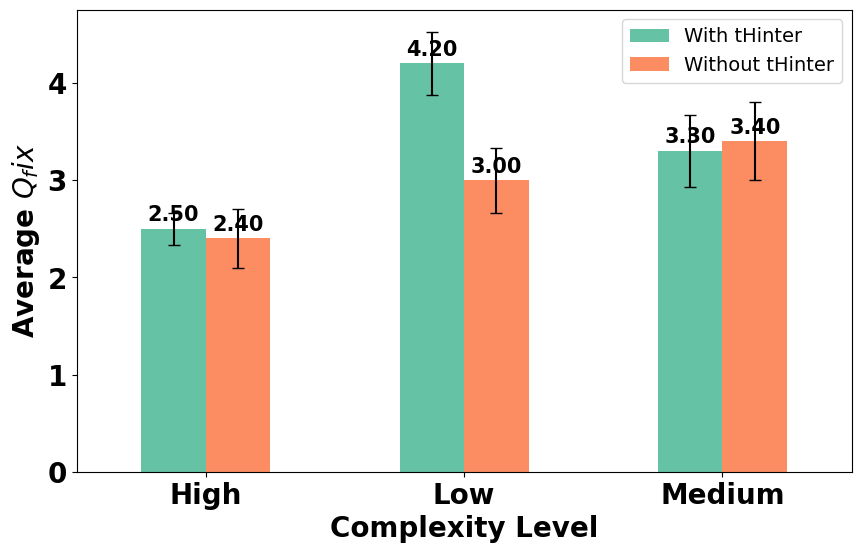}
        \vspace{-4pt}
        \label{fig:figure7}
        \end{minipage}
    }
    
    \vspace{-12pt}
    \caption{ Comparing of $Q_{fix}$ with/without \MIG during debugging by the LLM.}
    \label{fig:combined_figures3}
    \vspace{-8pt}
\end{figure*}

According to Figure~\ref{fig:figure6} on average a 13\% higher $Q_{fix}$ is achieved with \MIG. This indicates \MIG increases the perceived quality of changes made by the LLM during debugging by 13\%. \textit{Low} complexity code snippets benefit from \MIG most (a 40\% rise of $Q_{fix}$). One possible reason lies in the usually shorter and simpler structure of \textit{Low} complexity code snippets. \MIG facilitates the inference of the LLM by providing richer context information. Hence the translation errors are more easily noticed in a shorter and simpler structure and fixed by the LLM.

\begin{center}
\fbox{
    \parbox{0.9\linewidth}{\textbf{Summary:} 
    \MIG increases the likelihood of the LLM fixing translation errors in a single query by 59\% and enhances the quality of changes made during debugging by 13\%. By providing localized translation errors—essentially context information—the LLM is better equipped to recognize and correct translation errors.
    % 解释原因
    }
}
\end{center}
\subsection{Performance with Different Numbers of Test Cases Executed (RQ3)}
\textbf{Baselines.} We evaluate the key component of \MIG, the test case generation, by comparing the performance of \MIG with different numbers of test cases.
\begin{itemize}
    \item \textbf{10000 test cases:} All test cases generated are executed. Due to consideration of execution time, for those with over 10000 test cases, we set a cutoff of 10000.
    \item \textbf{200 test cases:} 200 test cases are randomly sampled from all test cases generated by \MIG and executed.  
    \item \textbf{50 test cases:} 50 test cases are randomly sampled from all test cases generated by \MIG and executed. 
\end{itemize}
\textbf{Results.} Comprehensively, more test cases benefit the performance of \MIG. Metrics drop significantly when only 50 test cases are executed. More test cases, essentially a higher line coverage, explore the behavior of the translated code more comprehensively. This would result in pointing to translation errors from a finer granularity. This is also in line with why \textit{Low} complexity code achieves the highest $R_{reduc}$. We also notice that comprehensively $H_{perceived}$ achieves the highest with \textit{Medium} complexity level code, no matter how many test cases are executed. After a discussion of C1 and C2, a loose conclusion can be drawn that a clear code structure benefits the performance of \MIG. Appropriate abstraction and modularization, decoupling between functional modules, and a logical sequence of function calls contribute to this clarity. Such a well-structured code allows correctly translated lines to be more accurately excluded by \MIG, thereby increasing the precision of localizing translation errors. Writing code with a clear structure often requires guidance from design patterns, which are typically not considered in overly simple code. Meanwhile, overly complicated code generally leverages more complicated algorithms and structures, difficult to understand for both humans and automatic methods. As a result, code at the \textit{Medium} complexity level achieves the highest $H_{perceived}$.
\begin{table}[h!]
\centering
\caption{Performance of \MIG with Different Numbers of Test Cases Executed. $H_{per.}$ is short for $H_{perceived}$.}
\begin{tabular}{l|c c|c c|c c|c c}
\toprule
\multirow{2}{*}{\makecell{\textbf{N.Test} \\ \textbf{Cases}}} 
& \multicolumn{2}{c|}{\textbf{Overall}} 
& \multicolumn{2}{c|}{\textbf{Low}} 
& \multicolumn{2}{c|}{\textbf{Medium}} 
& \multicolumn{2}{c}{\textbf{High}} \\ 
% \cmidrule(r){2-7}
 & \makecell{\textbf{$R_{reduc}\uparrow$}} & \makecell{\textbf{$H_{per.}\uparrow$}}
 & \makecell{\textbf{$R_{reduc}\uparrow$}} & \makecell{\textbf{$H_{per.}\uparrow$}} 
 & \makecell{\textbf{$R_{reduc}\uparrow$}} & \makecell{\textbf{$H_{per.}\uparrow$}} 
 & \makecell{\textbf{$R_{reduc}\uparrow$}} & \makecell{\textbf{$H_{per.}\uparrow$}} \\ 
\midrule
10000 & 0.71 & 3.18 & 0.76 & \textbf{3.17} & 0.70 & 3.18 & 0.61 & 3.17 \\ 
200   & \textbf{0.73} & \textbf{3.19} & \textbf{0.77} & \textbf{3.17} & \textbf{0.73} & \textbf{3.20} & \textbf{0.63} & \textbf{3.18} \\ 
50    & 0.72 & 2.73 & 0.76 & 2.70 & 0.74 & 2.74 & 0.61 & 2.74 \\ 
\bottomrule
\end{tabular}
\end{table}

\begin{center}
\fbox{
    \parbox{0.9\linewidth}{\textbf{Summary:} 
    More test cases, essentially higher line coverage,  benefit the performance of \MIG. This illustrates the reasonableness of the coverage guided test case generation strategy in \MIG. This strategy allows \MIG to explore the behavior of the translated code more comprehensively and localize translation errors from a finer granularity.
    % 解释原因
    }
}
\end{center}
\section{Case Studies}
We summarize two typical types of translation errors and present how could \MIG help with the debugging process.
\subsection{Nonequivalent Logic Flow}
The Python solution to LeetCode question \textit{finding trios with the lowest degree} defines a function \textit{minTrioDegree}. This function first constructs an adjacency matrix \textit{g} and a degree list \textit{degree} to represent the graph and track the degree of each node. Then, it iterates over all possible trios (i, j, k) in the graph. If a trio is found (i.e., all three nodes are connected to each other), it calculates the degree of the trio and updates the minimum degree. Finally, it returns the minimum trio degree if a valid trio is found, or -1 if no trio exists. A chained comparison is used in the iteration process: \textbf{if g[i][k] == g[j][k] == 1:}.

In the C++ version, this chained comparison is translated into \textbf{if (g[i][k] == g[j][k] == 1)}, which causes altered functionalities. In Python, this chained comparison checks both conditions: g[i][k] == 1 and g[j][k] == 1. However, in C++ this expression is evaluated as two separate comparisons: first, g[i][k] == g[j][k], which results in a boolean value (true or false), and then this boolean value is compared to 1. This leads to incorrect behaviors because the second comparison (true == 1 or false == 1) is not equivalent to checking if both g[i][k] and g[j][k] are equal to 1. To fix this translation error in C++, the chained comparison must be split into two separate conditions: \textbf{if (g[i][k] == 1 \&\& g[j][k] == 1}).

\MIG directly pinpoints this translation error. After seeing the output of \MIG, the debugging developer immediately recognizes that the chained comparison has been incorrectly translated. Hence, no laborious line-to-line inspection is needed. This translation error is fixed within 3 minutes.
\subsection{Nonequivalent Data Type}
 The Python solution to LeetCode question \textit{finding the minimum spped} defines a function \textit{minSpeedOnTime} that calculates the minimum speed required to travel a series of distances within a given time limit. The function uses binary search to determine the smallest possible speed that allows for the completion of the journey within the specified time. The check method calculates the total time required to cover the distances at a given speed and checks if it meets the time constraint. 
 
 However, translation introduces a risk of integer overflow. In the C++ version, variables such as \textit{t }(representing time) and \textit{speed} (representing speed) are stored as \textbf{integers}, the same as in the Python solution. Following multiplication operations like \textit{t *= speed} and \textit{t * 100} can easily exceed the maximum value that an int can hold, leading to an overflow. This overflow results in incorrect calculations, causing the algorithmic logic to fail. Python automatically manages integer overflow by dynamically allocating more memory. This means that even if the values being calculated become very large, Python will still handle them correctly without causing overflow. To fix this, variables should be stored as larger data types, such as \textbf{long long}.

 \MIG highlights lines where multiplication operations are conducted with variable \textit{t } and variable \textit{speed}. This inspires the debugging developer to examine the data type of those variables and change the data type to \textbf{long long}.

\section{Threats to Validity}
Here we present factors hindering the performance of \MIG, both within (internal threats) and outside of (external threats) the method.

\textbf{Internal Threats.} Firstly, \MIG does not apply to auto-translated code from languages not supported by LLVM, such as JavaScript or Verilog. However, since LLVM supports a wide range of programming languages, including nearly all commonly used ones, we believe this limitation will not significantly affect \MIG's effectiveness and usefulness. 
 Secondly, \MIG relies on fuzzing to generate test cases, where the mutation incurs randomness. If the input is particularly complex, it may result in some test cases that do not fully satisfy the constraints, potentially leading to reduced effectiveness. The \textit{localize by Expertise} component is not affected by this limitation and can continue to provide useful suggestions for debugging.

 \textbf{External Threats. } Firstly, if the translated code is not valid but merely a string of special characters, it could, in rare cases, cause \MIG to crash. However, since such code typically fails to compile, \MIG would return a compiler error and remind developers to debug according to the error message. Secondly, \MIG's performance may vary across codes from different translation tools.
Further evaluations of \MIG are needed to understand whether and how the performance of \MIG changes with translation tools following different technical paths. We will explore this in future work.

\section{Related work}
\subsection{Differential Testing}
Differential testing focuses on multiple systems with similar functionalities~[\citeauthor{ DifferentialTestingSoftware1998}~\citeyear{DifferentialTestingSoftware1998}]. It is difficult to determine the correct output given an input without prior knowledge. Differential testing solves this challenge by comparing the outputs from multiple counterpart software and thus provides an oracle~[\citeauthor{DBLP:conf/sigsoft/EvansS07}~\citeyear{DBLP:conf/sigsoft/EvansS07}]. To conduct differential testing, an essential step is the construction of test cases. Test cases can be constructed by human force~[\citeauthor{janaAbusingFileProcessing2012}~\citeyear{janaAbusingFileProcessing2012}] and automatic techniques such as mutation~[\citeauthor{brubakerUsingFrankencertsAutomated2014}~\citeyear{brubakerUsingFrankencertsAutomated2014}]. Coverage-guided test case generation is also leveraged to construct test cases exploring the behavior of the software under test comprehensively~[\citeauthor{chenGuidedDifferentialTesting2015}~\citeyear{chenGuidedDifferentialTesting2015}]. 
Differential testing has been widely adopted in defects uncovering~[\citeauthor{DBLP:conf/uss/BrumleyCLN07}~\citeyear{DBLP:conf/uss/BrumleyCLN07};~\citeauthor{ DBLP:conf/osdi/CadarDE08}~\citeyear{DBLP:conf/osdi/CadarDE08};~\citeauthor{DBLP:conf/ccs/ChapmanE11}~\citeyear{DBLP:conf/ccs/ChapmanE11};~\citeauthor{DBLP:conf/pldi/SrivastavaBMS11}~\citeyear{DBLP:conf/pldi/SrivastavaBMS11};~\citeauthor{DBLP:conf/pldi/YangCER11}~\citeyear{DBLP:conf/pldi/YangCER11};~\citeauthor{argyrosSFADiffAutomatedEvasion2016}~\citeyear{argyrosSFADiffAutomatedEvasion2016};~\citeauthor{chenCoveragedirectedDifferentialTesting2016}~\citeyear{chenCoveragedirectedDifferentialTesting2016};~\citeauthor{zhangAutomatedTransplantationDifferential2017}~\citeyear{zhangAutomatedTransplantationDifferential2017}]. Besides software, differential testing can also target fundamental parts modern software build upon.  For example, an automated fuzzer and a differential oracle are used to discover transient execution vulnerabilities in the CPU~[\citeauthor{hurSpecDoctorDifferentialFuzz2022}~\citeyear{hurSpecDoctorDifferentialFuzz2022}]. Deep learning libraries that are essential to many intelligent software can also be tested by fuzzing of neural architecture and defects are found by reports of inconsistency of outputs~[\citeauthor{DBLP:conf/icse/GuLZ022}~\citeyear{DBLP:conf/icse/GuLZ022}]. As the pair of codes before and after the translation tool should present the exact same functionalities, they can be regarded as counterpart software. Hence, differential testing is applicable in finding the translation errors incurred by translation tools. 
\subsection{Debugging}
Human factors have always been playing important roles in software engineering tasks, including debugging~[\citeauthor{gannon1979human}~\citeyear{gannon1979human}]. Debugging is a human effort intensive task, usually taking over 50 \% of the time and effort of a whole project~[\citeauthor{brooks1974mythical}~\citeyear{brooks1974mythical}]. During debugging, developers run the code and perform tasks manually multiple times~[\citeauthor{agrawal1991towards}~\citeyear{agrawal1991towards}]. By these means, they collect information such as sequences of steps performed, histories of variable values, function call hierarchies~[\citeauthor{DBLP:conf/kbse/AugustonJU02}~\citeyear{DBLP:conf/kbse/AugustonJU02}]. For the challenging nature of debugging lies in acquiring deep knowledge of the code~[\citeauthor{zamfirExecutionSynthesisTechnique2010}~\citeyear{zamfirExecutionSynthesisTechnique2010}]. To facilitate debugging, an essential step is to understand the human behavior during it. Developer actions are tracked to explore their information collection process~[\citeauthor{DBLP:journals/tse/KoMCA06}~\citeyear{DBLP:journals/tse/KoMCA06}]. Eye-tracking techniques are adopted to illustrate how they comprehend codes and styles~[\citeauthor{DBLP:conf/csee/SharifM10}~\citeyear{DBLP:conf/csee/SharifM10};~\citeauthor{DBLP:journals/ese/BinkleyDLMMS13}~\citeyear{DBLP:journals/ese/BinkleyDLMMS13}]as well as conducting code summarization ~[\citeauthor{rodeghero2014improving}~\citeyear{rodeghero2014improving}]. It is found that novice developers would debug more efficiently with the eye-gazing track information of experienced developers ~[\citeauthor{DBLP:conf/icmi/SteinB04}~\citeyear{DBLP:conf/icmi/SteinB04}]. This might be because the gazing track information offers extra knowledge and suggestions. To avoid the trouble of manually debugging, automatic debugging approaches are proposed~[\citeauthor{shapiro1982algorithmic}~\citeyear{shapiro1982algorithmic};~\citeauthor{weiser1984program}~\citeyear{weiser1984program};~\citeauthor{DBLP:conf/pldi/FritzsonGKS91}~\citeyear{DBLP:conf/pldi/FritzsonGKS91};~\citeauthor{machadoMZoltarAutomaticDebugging2013}~\citeyear{machadoMZoltarAutomaticDebugging2013};~\citeauthor{DBLP:conf/issta/LouGLZZHZ20}~\citeyear{DBLP:conf/issta/LouGLZZHZ20}]. To name some of them as examples, automated identification of faulty constraints can significantly decrease development and maintenance efforts for variability models~[\citeauthor{DBLP:conf/icse/LeFU0GT21}~\citeyear{DBLP:conf/icse/LeFU0GT21}]. Fed-DNN-Debugger debugs a client model by a nonintrusive metadata capture module and automated neural network model debugging module~[\citeauthor{duan2023fed}~\citeyear{duan2023fed}]. MRDebug debugs the target code by locating the root cause, analyzing several executions of the test case, and a Delta Debugging technique isolating the data relevant to bug triggering~[\citeauthor{DBLP:journals/tse/MoranBRT24}~\citeyear{DBLP:journals/tse/MoranBRT24}]. LLM can also be adopted in automatic debugging~[\citeauthor{lee2024unified}~\citeyear{lee2024unified};~\citeauthor{DBLP:conf/icse/NamMHVM24}~\citeyear{DBLP:conf/icse/NamMHVM24}]. For example, Panda, inspired by how experienced developers debug, provides context information to LLMs and thus gets troubleshooting recommendations~[\citeauthor{DBLP:conf/cidr/SinghVKKNGK24}~\citeyear{DBLP:conf/cidr/SinghVKKNGK24}].
 However, with the help of so many automatic tools, debugging remains human-effort intensive. A study with spectrum-based fault localization assistance finds that it may even weaken the ability to debug~[\citeauthor{xieRevisitAutomaticDebugging2016}~\citeyear{xieRevisitAutomaticDebugging2016}].
\subsection{Code Translation}
Translating code of one language to another is of crucial importance to many software engineering tasks, like code migration~[\citeauthor{DBLP:journals/peerjpre/AggarwalSH15}~\citeyear{DBLP:journals/peerjpre/AggarwalSH15}]. To avoid the human-intensiveness of manual translation, translation tools are required. One typical methodology translation tools follow is machine translation. A neuron network is trained to generate the parallel code given the source code from another programming language~[\citeauthor{DBLP:conf/acl/KoehnHBCFBCSMZDBCH07}~\citeyear{DBLP:conf/acl/KoehnHBCFBCSMZDBCH07};~\citeauthor{DBLP:conf/sigsoft/NguyenNN13}~\citeyear{DBLP:conf/sigsoft/NguyenNN13};~\citeauthor{DBLP:conf/oopsla/KaraivanovRV14}~\citeyear{DBLP:conf/oopsla/KaraivanovRV14};~\citeauthor{DBLP:journals/peerjpre/AggarwalSH15}~\citeyear{DBLP:journals/peerjpre/AggarwalSH15};]. However, such translation tool needs parallel code during the training process, which is hard to access. Hence, unsupervised method is also proposed~[\citeauthor{DBLP:conf/nips/RoziereLCL20}~\citeyear{DBLP:conf/nips/RoziereLCL20}]. Most machine translation based translation tools rely on BLEU to measure the performance, which is more suitable for natural languages, hence, a new metric is also proposed~[\citeauthor{DBLP:conf/icse/EghbaliP22}~\citeyear{DBLP:conf/icse/EghbaliP22}]. LLMs can also be adopted in a variety of code intelligence tasks, including code translation~[\citeauthor{roziere2023code}~\citeyear{roziere2023code};~\citeauthor{DBLP:conf/iui/RossMHMW23}~\citeyear{DBLP:conf/iui/RossMHMW23};~\citeauthor{DBLP:conf/icse/NiuLNCGL23}~\citeyear{DBLP:conf/icse/NiuLNCGL23};~\citeauthor{DBLP:conf/icse/NiuLNCGL23}~\citeyear{DBLP:conf/icse/NiuLNCGL23};~\citeauthor{DBLP:journals/corr/abs-2401-00288}~\citeyear{DBLP:journals/corr/abs-2401-00288};], although translations by LLM can be faulty~[\citeauthor{pan2024lost}~\citeyear{pan2024lost}]. Evaluating the similarity of a code pair is essential to translation tools. To evaluate the similarity between a code pair from the same language, text based comparation is typically used~[\citeauthor{DBLP:conf/wcre/Baker95}~\citeyear{DBLP:conf/wcre/Baker95};~\citeauthor{DBLP:journals/tse/KamiyaKI02}~\citeyear{DBLP:journals/tse/KamiyaKI02};~\citeauthor{DBLP:journals/tse/LiLMZ06}~\citeyear{DBLP:journals/tse/LiLMZ06};]. In a cross language context, both static and dynamic techniques are used~[\citeauthor{DBLP:conf/kbse/NafiKRRS19}~\citeyear{DBLP:conf/kbse/NafiKRRS19};~\citeauthor{DBLP:conf/sigsoft/MathewS21}~\citeyear{DBLP:conf/sigsoft/MathewS21}].
\section{Conclusion}
In this paper, we present \MIG, an automatic approach to localize translation errors in auto-translated code. It adopts differential testing and the core idea is to pinpoint lines in the translated code causing output differences with the source code. To avoid missing translation errors, \MIG leverages a fuzzing-based approach to generate valid test cases exploring the behavior of the translated code comprehensively. Pass or fail results in the differential testing only indicate the existence of translation errors instead of the exact location. To conduct localization from execution results, a heuristics-based localization algorithm is designed, leveraging both statistics and developers's expertise. \MIG is proven to be effective in both manual debugging and debugging with the LLM. Developers generally consider \MIG helpful and satisfactory. To inspire future work, we open source \MIG.

\bibliographystyle{ACM-Reference-Format}
\bibliography{main}

\end{document}